\documentclass[twocolumn,preprintnumbers,nofootinbib,superscriptaddress]{revtex4}
\usepackage{amsmath} \usepackage{graphicx} \usepackage{amsfonts}
\usepackage{array} \usepackage{amsthm} \usepackage{bm} %\usepackage{mathptmx}

\usepackage{latexsym}

\usepackage[breaklinks]{hyperref}
\usepackage{color}

\newcommand{\nn}{\nonumber}
\newcommand{\be}{\begin{equation}}
\newcommand{\ee}{\end{equation}}
\newcommand{\ba}{\begin{eqnarray}}
\newcommand{\ea}{\end{eqnarray}}
\newcommand{\bal}{\begin{align}}
\newcommand{\eal}{\end{align}}

\newcommand{\e}{{\rm e}}
\newcommand{\dd}{{\rm d}}

\newcommand{\bb}{\bibitem}

\newcommand{\al}{\alpha}
\newcommand{\la}{\lambda}
\newcommand{\La}{\Lambda}
\newcommand{\bt}{\beta}

\newcommand{\ga}{\gamma}

\newcommand{\De}{\Delta}

\newcommand{\de}{\delta}

\newcommand{\bw}{\begin{widetext}}
\newcommand{\ew}{\end{widetext}}

\def\abh{black hole }
\def\bh{black holes }
\def\aBH{black hole}
\def\BH{black holes}
\def\RN{Reissner-Nordstr\"om }

\begin{document}
%\begin{flushleft}???
%\end{flushleft}
%\begin{flushleft}???
%\end{flushleft}
\title{Phantom black holes and critical phenomena}

\author{Mustapha Azreg-A\"{\i}nou}\email{azreg@baskent.edu.tr}
\affiliation{Ba\c{s}kent University, Engineering Faculty, Ba\u{g}l\i ca Campus, Ankara, Turkey}
\author{Glauber T. Marques}\email{gtadaiesky@hotmail.com}
\affiliation{Universidade Federal Rural da Amaz\^{o}nia
ICIBE - LASIC, Av. Presidente Tancredo Neves 2501
CEP 66077-901 - Bel\'em/PA, Brazil}
\author{Manuel E. Rodrigues}\email{esialg@gmail.com}
\affiliation{Faculdade de Ci\^{e}ncias Exatas e Tecnologia, Universidade Federal do Par\'{a} - Campus Universit\'{a}rio de Abaetetuba, CEP 68440-000, Abaetetuba, Par\'{a}, Brazil}

%\date{}

\begin{abstract}
We consider the two classes cosh and sinh of normal and phantom black holes of Einstein-Maxwell-dilaton theory. The thermodynamics of these holes is characterized by heat capacities that may have both signs depending on the parameters of the theory. Leaving aside the normal Reissner-Nordstr\"om black hole, it is shown that only some phantom black holes of both classes exhibit critical phenomena. The two classes share a nonextremality, but special, critical point where the transition is continuous and the heat capacity, at constant charge, changes sign with an infinite discontinuity. This point yields a classification scheme for critical points. It is concluded that the two unstable and stable phases coexist on one side of the criticality state and disappear on the other side, that is, there is no configuration where only one phase exists. The sinh class has an extremality critical point where the entropy diverges. The transition from extremality to nonextremality with the charge held constant is accompanied by a loss of mass and an increase in the temperature. A special case of this transition is when the hole is isolated (microcanonical ensemble), it will evolve by emission of energy, which results in a decrease of its mass, to the final state of minimum mass and vanishing heat capacity. The Ehrenfest scheme of classification is inaccurate in this case but the generalized one due to Hilfer leads to conclude that the transition is of order less than unity. Fluctuations near criticality are also investigated.

\end{abstract}


\maketitle

\section{Introduction\label{sec1}}

Scaling laws are important tools in modern engineering and science. As is well known, some mathematical theorems and useful physical formulas were derived employing the methods of scaling theory or dimensional analysis~\cite{book2}. In thermodynamics, scaling laws were discovered within theories of critical phenomena, which started to emerge due to observations of extremely large fluctuations at criticality in 1869~\cite{book3}.

In 1933 Ehrenfest introduced the first classification of phase transitions in classical thermodynamics based on finite jumps in the derivatives of the free energy, followed in 1938 by that of Landau and Lifshitz based on symmetry considerations~\cite{hist}. According to Ehrenfest classification, the order of the transition is the positive integer $p$ such that all derivatives, of a given potential, of order $\leq p-1$ are continuous while the $p$\textsuperscript{th} (partial) derivatives have finite jump discontinuities. Both classifications were later found to be inaccurate methods of classifying phase transitions due to the existence of infinite jump discontinuities (divergencies) in the heat capacity of some transitions and the occurrence of real (noninteger) orders of transition~\cite{Hilfer,Hilfer2}.

In the 1970s, a modern, rather simplified, scheme of classification was adopted by which the transitions are binary, either of first order, where the system absorbs some amount of heat at constant temperature, or continuous, that is, connected with a continuous change of symmetry.

Other classifications have continued to emerge~\cite{Hilfer}, concluding that transitions (1) with \emph{real} $p>1$ are continuous, (2) with $p<1$, called anequilibrium transitions~\cite{Hilfer2}, correspond to nonequilibrium thermodynamics and are characterized by an infinite entropy at the critical point (CP), and (3) those with $p=1$, first order transitions, are marginal.

Phase transitions and critical phenomena in \abh thermodynamics continue to occupy a large part of the \abh literature~\cite{Son:2012vj}-\cite{AdS}. Most of these investigations focused on anti-de Sitter (AdS) \BH~\cite{Wu}-\cite{AdS} due, on the one hand, to the existing similarities in their phase transitions with those of both the magnetic and the van der Waals liquid-gas systems~\cite{book}. On the other hand, the extension of applicability of the principal of AdS/CFT correspondence~\cite{Maldacena1}, to phenomena described by fluid dynamics and other systems and to provide simple holographic dual descriptions to almost all CFT phenomena (including their phase transitions)~\cite{Maldacena2}-\cite{Maldacena6}, has motivated such investigations.

Failure to observe a van der Waals liquid-gas like phase transition in some AdS \BH, in that the corresponding $PV$-diagram does not show a point of inflection at the CP, was the case in~\cite{Wei}. Very recently, a similar failure has been noticed for \bh with torsion~\cite{Ma}. These new types of phase transitions, which make the case in this work too, will lead us to introduce a precise classification of (\aBH) CP's, which is applicable to all types of transitions.

The presence of cosmological phantom fields continues to receive supports from both collected
observational data~\cite{Koma} and theoretical models~\cite{dyn}. All these programs have pointed out an accelerated expansion of the universe, dominated by an exotic fluid of negative pressure. Furthermore, there are
evidences suggesting the exotic fluid could be of phantom nature~\cite{nat1,nat2}. Since then, an interest in phantom fields has grown and resulted in many phantom \abh solutions~\cite{gr}-\cite{gerard2}. In recent years, many issues pertaining to phantom \BH, such as thermodynamic stability~\cite{thermo} and light paths~\cite{light,light2} have been dealt with. It is within the spirit of the above-mentioned motivations that we investigate the nature of the phase transitions and the CP's of some of the phantom \bh derived in~\cite{gerard1}.

In this work, attention is given to the so-called cosh and sinh phantom \bh of Einstein-Maxwell-dilaton theory (EMD)~\cite{gerard1}. The action governing the dynamics of these solutions depends on two discreet numbers $\eta_1=\pm 1$ and $\eta_2=\pm 1$, which determine the nature (normal or phantom) of the couplings of the dilaton and electromagnetic fields, respectively, and on a continuous parameter $\la$, which is the real dilaton-Maxwell coupling constant. Among the conclusions we reach in this work, the existence of a one nonextremality CP for the cosh solutions and two, nonextremality and extremality, CP's for the sinh ones, with the explicit dependence of the CP's on ($\la,\eta_1,\eta_2$). As to the critical exponents of the various thermodynamic variables, they do not depend on ($\la,\eta_1,\eta_2$) if the CP is a nonextremality one and they do depend on ($\la,\eta_1,\eta_2$) in the case of the extremality CP. Various diagrams, corresponding to a $PV$-diagram (or to magnetic field-magnetization diagram), do not exhibit points of inflection at criticality. This reveals a new type of transitions for these \BH, which were encountered for other \abh solutions, as mentioned above, but never distinguished from known transitions; We will achieve that on introducing a classification of CP's.

In Ref.~\cite{Cai:1997cv}, the thermodynamics of only normal (non-phantom) sinh \bh has been investigated. As we shall see below, normal sinh \bh have two horizons the interior of which is singular for all $\la \neq 0$ (in the notation of~\cite{Cai:1997cv}, $a^2=\la^2$). The case $\la=0$ is \RN \abh which has a regular inner horizon. In this work we will exclude from our analysis all sinh \bh having a singular inner horizon for at extremality, where the two horizons coincide and both become singular, the thermodynamics becomes subtle.

In Sect.~\ref{sec2} we introduce the action governing the dynamics of EMD theory, the metric of the cosh and sinh \BH, and set the conditions for not having a singular inner horizon. In Sect.~\ref{sec3}, unless otherwise specified, we restrict ourselves to canonical ensembles (fixed-charge ensembles), review and derive the thermodynamics of these \BH. Sect.~\ref{sec4} is devoted to the critical behavior of the different thermodynamic functions and variables of these \BH.

In Sect.~\ref{sec5} we distinguish two types of CP's, one nonextremality CP shared by both cosh and sinh \bh and the other, called extremaliy CP, emerges near extremality conditions of the sinh solutions. We introduce a classification of CP's which distinguishes $n$th order (hyper)surface CP's and $n$th order discreet ones. For the nonextremality CP, the classification due to Ehrenfest yields a second order phase transition at it. For the extremaliy CP, Ehrenfest classification fails since in this case the entropy diverges at criticality (the derivatives of all orders of the free energy diverge!). However, recalling the classification due to Hilfer~\cite{Hilfer,Hilfer2}, we argue that the transition at the extremaliy CP is anequilibrium of order less than unity: $p<1$. The critical exponents for the nonextremality CP obey the usual scaling laws and those related to the extremaliy CP obey similar scaling laws derived from the usual ones by mere translations.

In Sect.~\ref{secF} we restrict ourselves to canonical and microcanonical ensembles. For the nonextremality CP, it is shown that the moment of the mass fluctuation diverges near criticality if the \abh is in contact with a heat bath with which it exchanges heat only. In contrast, for the extremality CP, it is shown that the relative root mean square dispersions vanish as criticality is approached, which yields no breakdown in thermodynamics. We conclude in~\ref{secC}.

\section{Normal phantom EMD \bh \label{sec2}}

The action for EMD theory with phantom Maxwell and/or dilaton field reads
\begin{equation}\label{2.1}
S=-\int \dd ^{4}x \sqrt{-g}\,\;[R-2\eta_1 g^{\mu\nu}
\partial_{\mu}\varphi\partial_{\nu}\varphi +\eta_2 \e^{2\lambda\varphi}
F_{\mu\nu}F^{\mu\nu}]\,,
\end{equation}
where $\lambda$ is the
real dilaton-Maxwell coupling constant, and $\eta_1=\pm 1$, $\eta_2=\pm 1$. Normal EMD
corresponds to $\eta_2=\eta_1=+1$, while phantom couplings of the dilaton field $\varphi$
or/and Maxwell field $F = \dd A$ are obtained for $\eta_1=-1$ or/and $\eta_2=-1$.

The metrics of the so-called cosh and sinh solutions, derived in~\cite{gerard1}, take the form
\begin{align}
& \dd s^{2}=f_{+}f_{-}^{\gamma}\dd t^{2}
-f_{+}^{-1}f_{-}^{-\gamma}\dd r^{2}
-r^{2}f_{-}^{1-\gamma}\dd \Omega^{2}\nn\\
\label{2.2}& F=-\frac{Q}{r^2}\,\dd r\wedge \dd t\, ,\quad
\e^{-2\lambda\varphi}=f_{-}^{1-\gamma}\\
&f_{\pm}=1- \frac{r_{\pm}}{r}\,,\quad \ga=\frac{1-\eta_1\la^2}{1+\eta_1\la^2} \nn\\
\label{2.3}&\frac{\eta_2}{1+\ga}<0 \text{ for cosh}\,,\quad \frac{\eta_2}{1+\ga}>0 \text{ for sinh}\\
& \ga \in (-\infty,-1)\cup[1,+\infty) \text{ if } \eta_{1}=-1\nn\\
\label{2.4}&\ga \in (-1,+1] \text{ if } \eta_{1}=+1
\end{align}
where we have introduced the parameter $\ga$ following the notation of~\cite{thermo,light}. These are asymptotically flat spherically symmetric black holes of mass $M$, electric charge $Q$ and event horizon $r_{+}>0$ related by~\cite{gerard1}
\begin{equation}\label{2.6}
2M  =r_{+}+\gamma r_{-}\,,\quad 2Q^2=\eta_{2}(1+\gamma)r_{+}r_{-}
\end{equation}
Since $Q$ is real, $r_-$ and $\eta_2(1+\ga)$ must have the same sign. Using this fact in~\eqref{2.3}, we have $r_-<0$ for the cosh solution and $r_->0$ for the sinh one. The case $\ga =1$ ($\lambda =0$) corresponds to normal \RN \abh if $\eta_2=+1$ or to phantom \RN \abh if $\eta_2=-1$ ($\eta_1$ is undefined in this case for there is no scalar field).

As we shall see in the following two sections, the thermodynamics and critical phenomena of the sinh solutions are more involved than those of the cosh solutions. For the latter solutions $r_+>0$ and $r_-<0$, so that $r_+$ never approaches $r_-$. Hence, the cosh solutions do not show an extremality limit and their temperature is always positive [Eq.~\eqref{3.1}]. Under which conditions do the sinh solutions exhibit extremality? The curvature scalar of the metric~\eqref{2.2} near $r_-$ ($r_+$ is a regular horizon and all the scalar invariants are finite there) is
\begin{equation}\label{2.7}
R=\frac{(\gamma ^2-1) r_-^2 (r-r_+)}{2 r^{3+\gamma } (r-r_-)^{2-\gamma }}
\end{equation}
which is singular if $\ga <2$. As the thermodynamics of singular horizons is not defined, extremality in this case ($r_+\to r_-$) is a subtle issue. We will drop the case where $R$ diverges at $r_-$ and consider the more restricted domain\footnote{In the case of extremal \bh $r_+\equiv r_-$, besides the interval~\eqref{2.8}, $R$ is finite in the limit $r\to r_+\equiv r_-$ in the interval $1< \ga <2$ too; however we won't include this interval in our discussion for extreme \bh are not stable. Moreover, extreme \bh may have nonthermal spectrum or ill-defined temperature~\cite{spec,spec2}.}
\begin{equation}\label{2.8}
    \ga =1\;\;\text{or}\;\;\ga \geq 2\quad (\text{for sinh})
\end{equation}
yielding $\eta_2=1$ for sinh solutions. As we mentioned earlier, $\eta_1$ is arbitrary for $\ga =1$ as there is no scalar field. For $\ga\geq 2$, $\eta_1=-1$ by\eqref{2.4}. Thus, except the known case $\ga =1$ corresponding to normal \RN \aBH, only phantom EManti-D sinh \bh have regular inner horizons.

\section{Thermodynamics \label{sec3}}

From now on we consider only positive charges: $Q>0$. More on the thermodynamics of these \bh is found in~\cite{thermo}. In this work, unless otherwise specified, we restrict ourselves to canonical ensembles (fixed-charge ensembles), where the \abh is in contact with a heat bath with which it exchanges heat only, and we focus on the critical behavior of these \BH.

The temperature of EMD \bh does not depend on the way the scalar field $\varphi$ is coupled to gravity (minimal or conformal coupling) and its mathematical expression is still given by Hawking formula
\begin{equation}\label{3.1}
    T=\frac{\partial_r(f_{+}f_{-}^{\gamma})/\partial r}{4\pi}\Big|_{r_+}=\frac{(r_+-r_-)^{\ga}}{4\pi r_+{}^{1+\ga}}.
\end{equation}
The entropy does generally depend on how the scalar field couples to gravity. In the case of the action~\eqref{2.1}, since the scalar field is minimally coupled to gravity, the expression of the entropy is not altered~\cite{C1}-\cite{C6} and is still given by Hawking formula
\begin{equation}\label{3.2}
    4S=\text{area of the horizon}=4\pi r_+{}^{1+\ga}(r_+-r_-)^{1-\ga}.
\end{equation}
In these expressions of $T$ and $S$ the only occurring parameters are $\ga$ and $r_-$ ($r_+$ is taken as a variable), but the latter depend via~\eqref{2.2} to~\eqref{2.6} on the other parameters of the theory. We thus expect different thermodynamical critical phenomena for the four spices of the EMD theory: normal EMD ($\eta_1=1$, $\eta_2=1$), Eanti-MD ($\eta_1=1$, $\eta_2=-1$), EManti-D ($\eta_1=-1$, $\eta_2=1$), and Eanti-Manti-D ($\eta_1=-1$, $\eta_2=-1$).

Using the second equation in~\eqref{2.6}, we first express $r_-$ in terms of ($r_+,Q$) then substitute into~\eqref{3.2}, the first equation in~\eqref{2.6}, and~\eqref{3.1} to derive $S$, $M$, and $T$ in terms of ($r_+,Q$)
\begin{align}
\label{3.3}&M=\frac{r_+}{2}+\frac{\eta_2\ga Q^2}{(1+\ga)r_+}\\
\label{3.4}&S=\frac{\pi r_+{}^{2\ga}}{(1+\ga)^{1-\ga}}[(1+\ga)r_+{}^2-2\eta_2 Q^2]^{1-\ga}\\
\label{3.5}&T=\frac{[(1+\ga)r_+{}^2-2\eta_2 Q^2]^{\ga}}{4\pi (1+\ga)^{\ga}r_+{}^{1+2\ga}}.
\end{align}
Let
\begin{equation}\label{est1}
A\equiv Q/r_+
\end{equation}
denotes the value of the electric potential $A_0$ on the horizon. We derive the first law of thermodynamics for these \BH, $\dd M=(\partial M(S,Q)/\partial S)_Q\dd S+(\partial M(S,Q)/\partial Q)_S\dd Q$, on evaluating the two derivatives $(\partial M(S,Q)/\partial S)_Q$ and $(\partial M(S,Q)/\partial Q)_S$ by
\begin{align}
\label{3.5a}&\Big(\frac{\partial M(S,Q)}{\partial S}\Big)_Q=\frac{(\partial M(r_+,Q)/\partial {r_+})_Q}{(\partial S(r_+,Q)/\partial {r_+})_Q}\\
&\Big(\frac{\partial M(S,Q)}{\partial Q}\Big)_S=-\frac{(\partial S(r_+,Q)/\partial Q)_{r_+}}{(\partial S(r_+,Q)/\partial {r_+})_Q}\Big(\frac{\partial M(r_+,Q)}{\partial r_+}\Big)_{Q}\nn\\
\label{3.5b}& \quad \quad \quad \quad\quad\quad\quad+\Big(\frac{\partial M(r_+,Q)}{\partial Q}\Big)_{r_+}.
\end{align}
Using~\eqref{3.3} and~\eqref{3.4} along with~\eqref{3.5} and the definition of $A$ [Eq.~\eqref{est1}], it is straightforward to show that the right-hand sides (r.h.s) in~\eqref{3.5a} and~\eqref{3.5b} are $T$ and $\eta_2A$, respectively. We have thus established~\cite{thermo}
\begin{equation}\label{est}
\dd M=T\dd S+\eta_2A\dd Q.
\end{equation}

\begin{figure}[h]
\centering
  \includegraphics[width=0.45\textwidth]{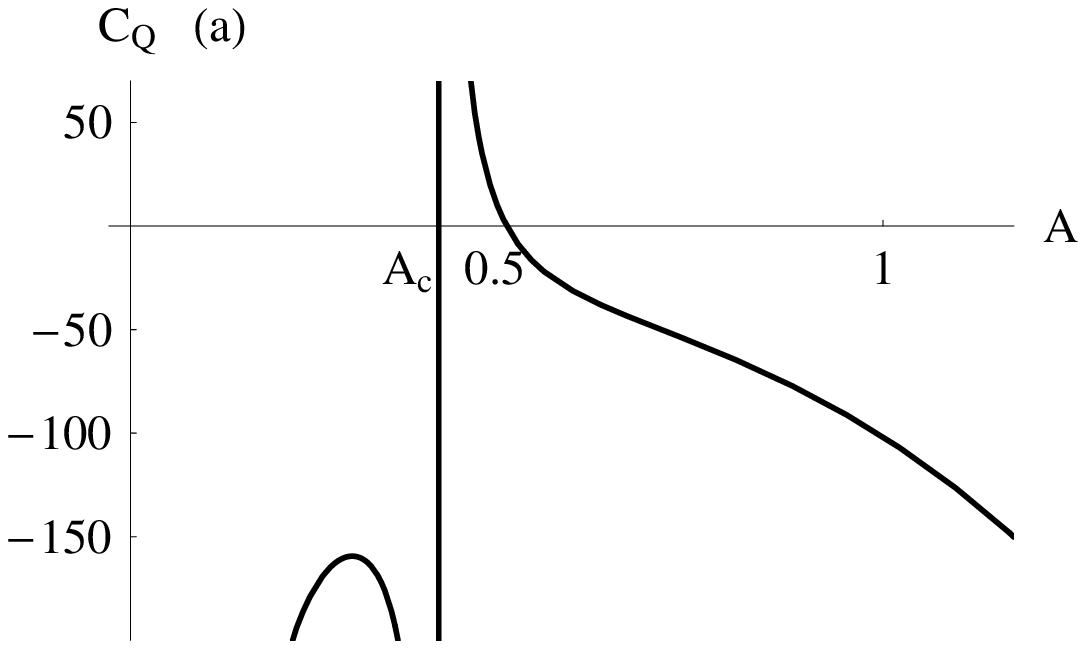} \includegraphics[width=0.45\textwidth]{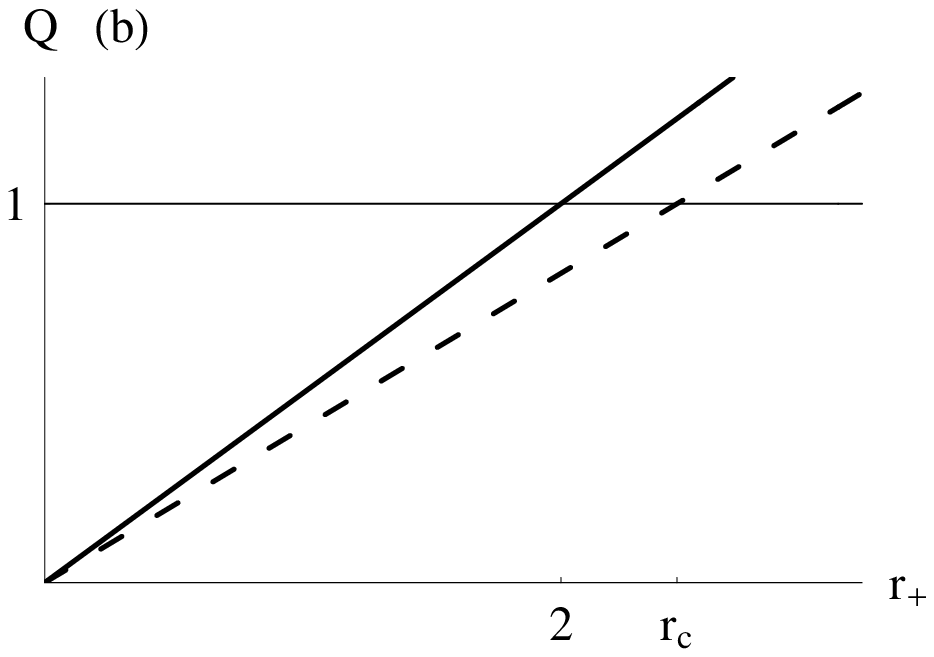}\\
  \caption{\footnotesize{The two plots are sketched for $\ga =-2$, corresponding to the cosh solution of EManti-D [Eq.~\eqref{3.9}] yielding $A_c=1/\sqrt{6}\simeq 0.408$. We have chosen $Q_c=1$ leading to $T_c=(1/\pi)\sqrt{3^3/2^{13}}\simeq 0.018$ and $S_c=128\pi/9\simeq 44.68$. (a): Plot of $C_Q(A)$ at constant $Q=Q_c$. The region $A<A_c$ is certainly unstable. The EManti-D \abh} solution undergoes a phase transition or change of degree of stability at $A=\sqrt{\eta_2(1+\ga)/(2\ga)}=1/2$. (b): Plot showing in the $r_+$-$Q$ plane the lines $L_1$ and $L_2$ [Eq.~\eqref{12bis}] where $C_Q$ is zero (continuous line) or diverges (dashed lines), respectively. There is no line that represents extremality and the \abh region is the whole first quadrant. As one moves on a line of constant charge from the left to the right, $r_+$ increases and $A$ decreases. The stable region is sandwiched between the continuous and the dashed lines.}\label{Fig1}
\end{figure}

\begin{figure}[!htb]
\centering
  \includegraphics[width=0.45\textwidth]{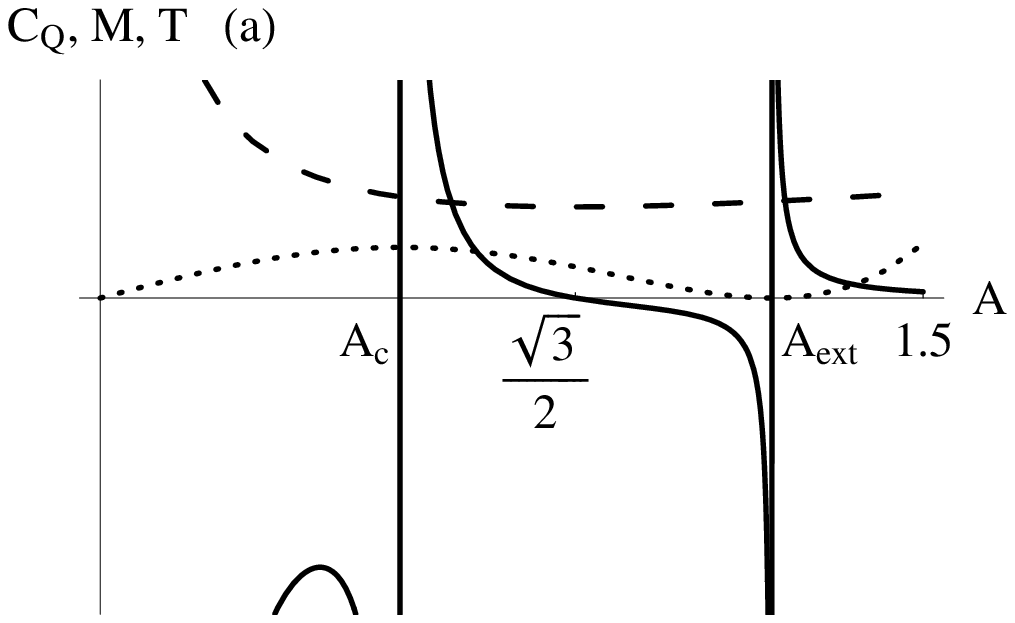} \includegraphics[width=0.45\textwidth]{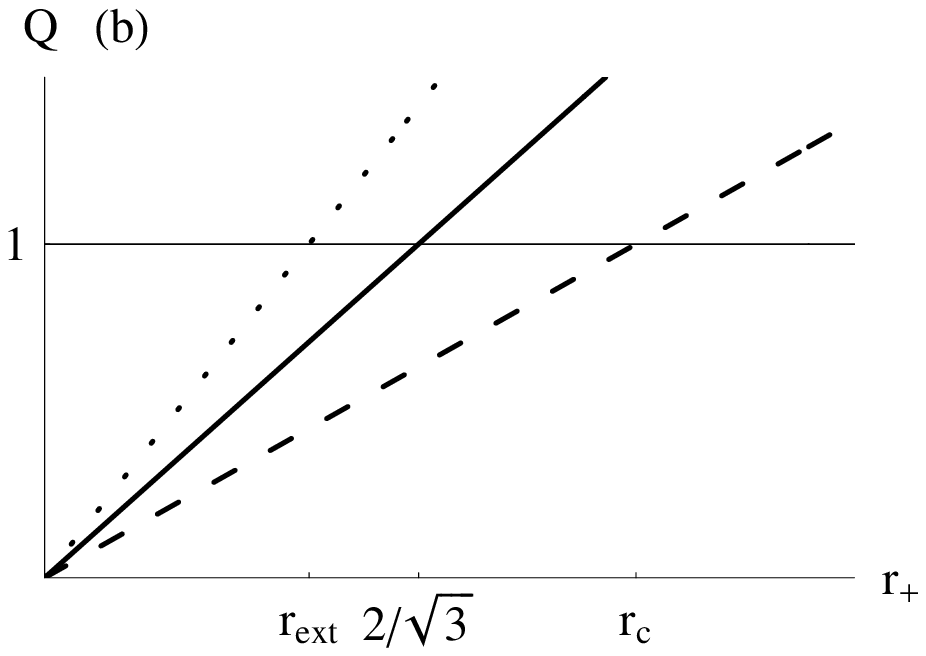}\\
  \caption{\footnotesize{The plots are sketched for $\ga =2$, corresponding to the sinh solution of EManti-D [Eq.~\eqref{3.12}] yielding $A_{c}=\sqrt{3/10}\simeq 0.548$. We have chosen $Q_c=Q_{\text{ext}}=1$ leading to $A_{\text{ext}}\equiv Q_{\text{ext}}/r_{\text{ext}}=\sqrt{3/2}\simeq 1.225$ and $T_c=2\sqrt{6/5}/(25\pi)\simeq 0.028$. (a): Plots of $C_Q(A)$ (continuous line), $M(A)$ (dashed line), and $T(A)$ (dotted line) all at constant $Q=Q_c$. The mass has its minimum value at the point where $C_Q=0$ [see~\eqref{3.7}, \eqref{3.7bb}] and the temperature has its extrema at ($A_c,A_{\text{ext}}$) [see~\eqref{3.7}, \eqref{3.7bb}]. The \abh region corresponds to $A<A_{\text{ext}}$ so one ignores the branch of the plot for $A>A_{\text{ext}}$. The region $A<A_{c}$ is certainly unstable. The EManti-D \abh solution undergoes a phase transition or change of degree of stability at $A=\sqrt{\eta_2(1+\ga)/(2\ga)}=\sqrt{3}/2$. There are two phase transitions one at $A_c$ (from a massive \abh to a lower mass one) and the other at $A_{\text{ext}}$ (from extremality to nonextremality). (b): Plot showing in the $r_+$-$Q$ plane the lines where $C_Q$ is zero [continuous line ($L_1$)] or diverges [dashed line ($L_2$) and dotted line). The \abh region is on the right of the dotted line, which represents extremality. The stable region is sandwiched between the continuous and the dashed lines.}}\label{Fig2}
\end{figure}

\begin{figure}[!htb]
\centering
  \includegraphics[width=0.45\textwidth]{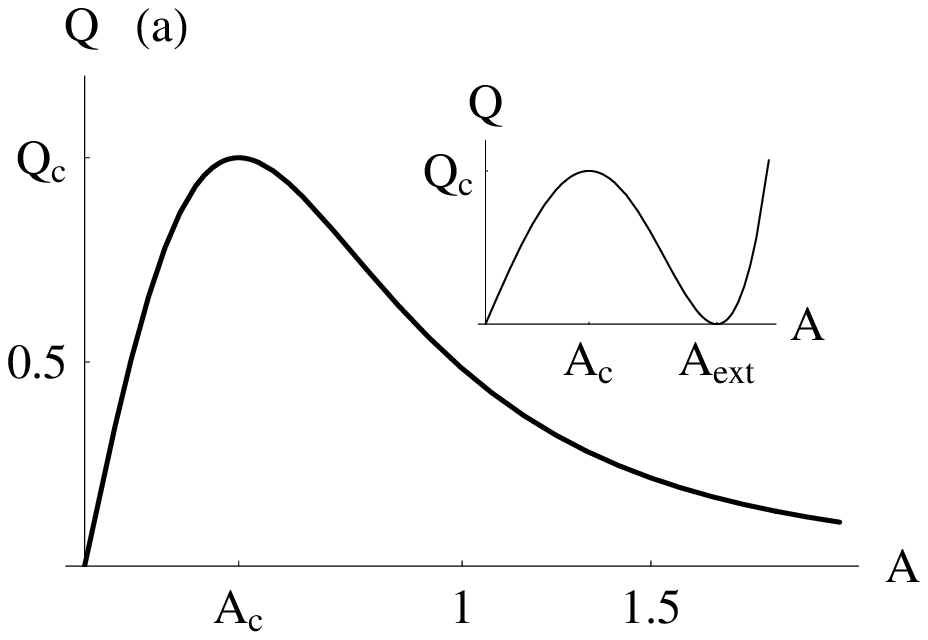} \includegraphics[width=0.45\textwidth]{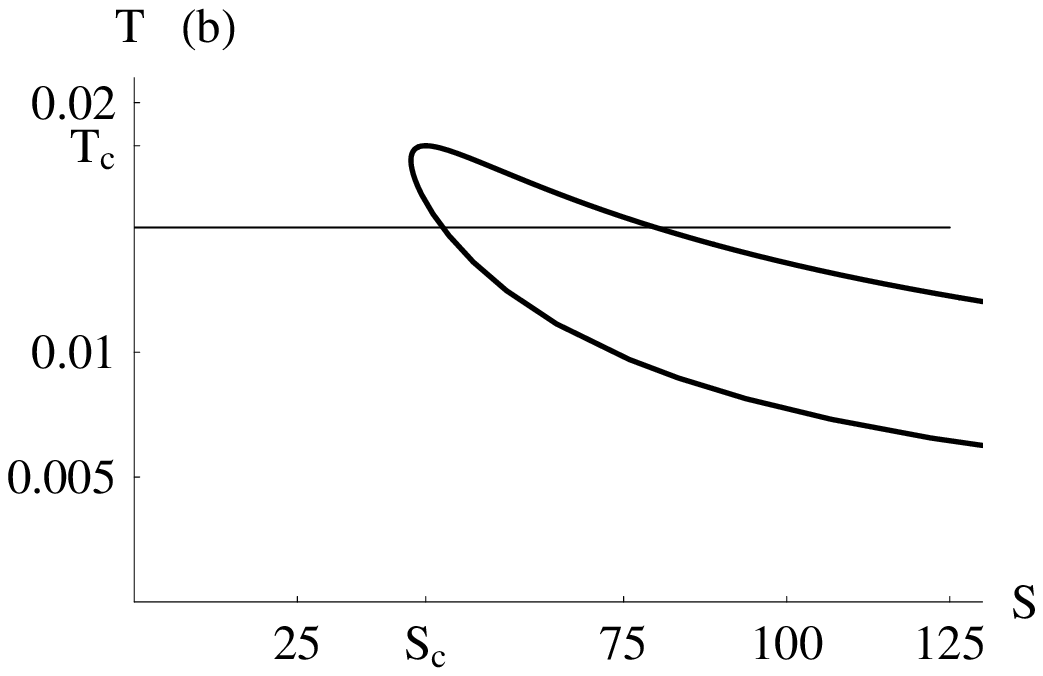} \includegraphics[width=0.45\textwidth]{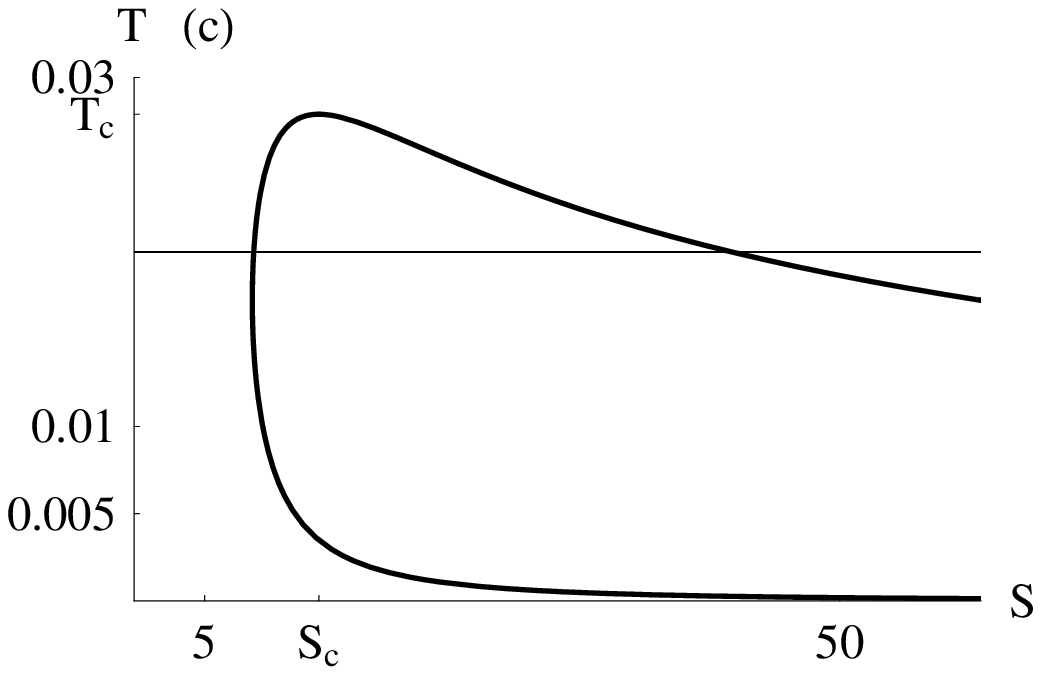}\\
  \caption{\footnotesize{The main plot (a) and the plot (b) are sketched for $\ga =-2$, corresponding to the cosh solution of EManti-D [Eq.~\eqref{3.9}] yielding $A_c=1/\sqrt{6}\simeq 0.408$. We have chosen $Q_c=1$ leading to $T_c=(1/\pi)\sqrt{3^3/2^{13}}\simeq 0.018$ and $S_c=128\pi/9\simeq 44.68$. The mini plot (a) and the plot (c) are sketched for $\ga =2$, corresponding to the sinh solution of EManti-D [Eq.~\eqref{3.12}] yielding $A_{c}=\sqrt{3/10}\simeq 0.548$. We have chosen $Q_c=Q_{\text{ext}}=1$ leading to $A_{\text{ext}}\equiv Q_{\text{ext}}/r_{\text{ext}}=\sqrt{3/2}\simeq 1.225$, $T_c=2\sqrt{6/5}/(25\pi)\simeq 0.028$ as in Fig.~\ref{Fig2} (a) and $S_c=25\pi/6\simeq 13.09$. (a): Plots of $Q(A)$ at constant $T=T_c$ for the cosh (main plot) and sinh (mini plot) solutions of EManti-D. The region $A<A_c$ is certainly unstable. For the plot of sinh, as $T_c\to T_{\text{ext}}=0$, the crest moves to infinity and the curve splits into the two vertical lines $A=0$ and $A=A_{\text{ext}}$, which are solutions of $F(A)=0$. On either line, the charge is undefined. (b): Parametric plot of $T(S)$ at constant $Q=Q_c$ for the cosh solution of EManti-D [Eq.~\eqref{3.9}]. No point of inflection at the point ($S_c,T_c$) nor a point of inflection at the point ($T_c,S_c$) for the $S(T)$ plot. The upper branch represents the unstable phase. (c): Parametric plot of $T(S)$ at constant $Q=Q_c$ for the sinh solution of EManti-D [Eq.~\eqref{3.12}].}}\label{Fig3}
\end{figure}

\begin{figure}[!htb]
\centering
  \includegraphics[width=0.45\textwidth]{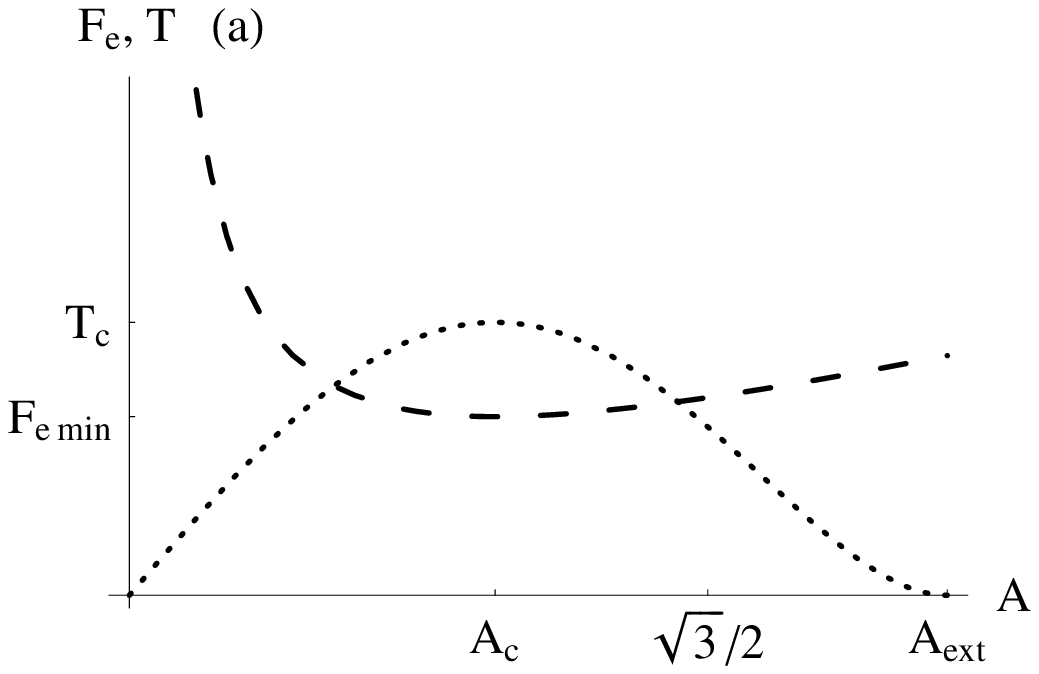} \includegraphics[width=0.45\textwidth]{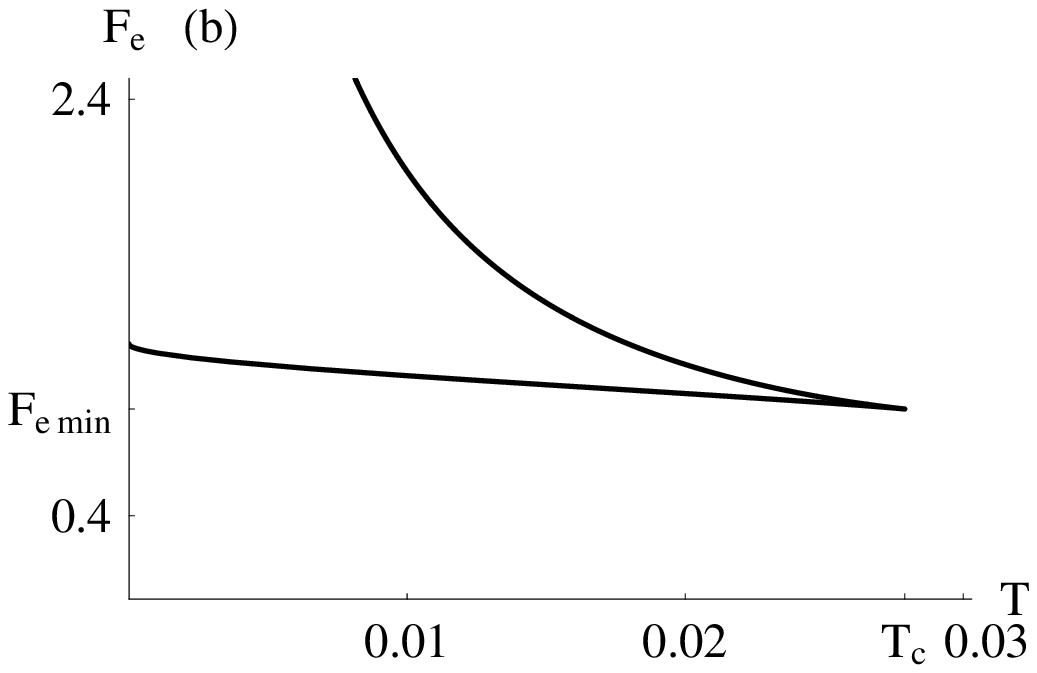}\\
  \caption{\footnotesize{The plots are sketched for $\ga =2$, corresponding to the sinh solution of EManti-D [Eq.~\eqref{3.12}] yielding $A_{c}=\sqrt{3/10}\simeq 0.548$. We have chosen $Q_c=Q_{\text{ext}}=1$ leading to $A_{\text{ext}}\equiv Q_{\text{ext}}/r_{\text{ext}}=\sqrt{3/2}\simeq 1.225$ and $T_c=2\sqrt{6/5}/(25\pi)\simeq 0.028$. (a): Plots of the free energy $F_{\text{e}}(A)$ (dashed line) and $T(A)$ (dotted line) at constant $Q=Q_c$ using different vertical scales. The two functions reach their extreme values $F_{\text{e}\,\text{min}}=Q_c/(2A_c)=\sqrt{5/6}\simeq 0.913$ [Eqs.~\eqref{3.7cc}, \eqref{3.13}] and $T_{\text{max}}=T_c$, respectively at $A=A_c$. (b): Parametric plot of $F_{\text{e}}(T)$ at constant $Q=Q_c$. The upper branch represents the unstable phase. The plots of this figure complete those of Fig.~\ref{Fig2}.}}\label{Fig4}
\end{figure}

The equation of state (EOS) is derived on substituting $r_-=2\eta_2 Q^2/[(1+\ga)r_+]$ and $r_+=Q/A$ into~\eqref{3.1}
\begin{equation}\label{3.6}
    4\pi (1+\ga)^{\ga}TQ=A(1+\ga-2\eta_2 A^2)^{\ga}.
\end{equation}
For subsequent use, we re-write the EOS as
\begin{equation}\label{3.6b}
    TQ=\frac{A(1+\ga-2\eta_2 A^2)^{\ga}}{4\pi (1+\ga)^{\ga}}\equiv F(A).
\end{equation}

\section{Critical behavior \label{sec4}}

By analogy with classical thermodynamics, the critical behaviors of thermodynamical systems occur at points (CP's) where, say, the heat capacities and generalized susceptibilities diverge~\cite{book3,book}. According to modern classification, this corresponds to a second order or continuous phase transition, where only the second order partial derivatives of the free energy have either a jump discontinuities or infinite ones but its first order partial derivatives are continuous. The heat capacity at constant charge, $C_Q\equiv T(\partial S/\partial T)_Q$, which, using~\eqref{3.4} and~\eqref{3.5}, reduces to
\begin{multline}\label{3.7}
    C_Q=T\frac{(\partial S(r_+,Q)/\partial r_+)_Q}{(\partial T(r_+,Q)/\partial r_+)_Q}=-2 \pi  r_+{}^{2 \gamma } (1+\gamma )^{\gamma -1}\times \\
    \frac{[(1+\gamma ) r_+{}^2-2 \eta _2 \gamma  Q^2] [(1+\gamma ) r_+{}^2-2 \eta _2
Q^2]^{1-\gamma }}{[(1+\gamma ) r_+{}^2-2 \eta _2 (1+2 \gamma ) Q^2]}.
\end{multline}
For all $\ga$, $C_Q$ diverges at the CP
\begin{equation}\label{3.8}
    r_c{}^2=\frac{2\eta _2 Q_c{}^2 (1+2 \gamma )}{1+\gamma }
\end{equation}
where $Q_c$ is the critical charge. If $A_c$ denotes the critical electric potential, combining $A_c=Q_c/r_c$ with~\eqref{3.8} we obtain
\begin{equation}\label{3.13}
    A_c{}^2=\frac{\eta_2(1+\ga)}{2(1+2\ga)}.
\end{equation}

The requirement $r_c{}^2> 0$ sets some constraints on $\ga$: From~\eqref{2.3} we see that $1+2\ga< 0$ ($\ga< -1/2$) for the cosh solutions and that $1+2\ga> 0$ ($\ga > -1/2$) for the sinh ones, which is already satisfied by~\eqref{2.8}. Now, by the other requirement, $r_c\geq r_-(=2\eta_2 Q_c{}^2/[(1+\ga)r_c])$, which is always satisfied for the cosh solutions, we must have $\ga\geq 0$ for the sinh solutions and this is again satisfied by~\eqref{2.8}. Finally, combining these conclusions with~\eqref{2.3}, \eqref{2.4}, and~\eqref{2.8}, we obtain the following conditions for having critical phenomena at $r_c$
\begin{align}
&\text{cosh}: \gamma < -1/2\nn\\
\label{3.9}&\gamma \in (-\infty ,-1)\to \eta _1=-1, \eta _2=+1\\
\label{3.10}&\gamma \in (-1,-1/2)\to \eta _1=+1, \eta _2=-1;\\
&\text{sinh}: \ga =1\;\;\text{or}\;\;\ga \geq 2\nn\\
\label{3.11}&\gamma =1\to \eta _1=\pm 1, \eta _2=+1\\
\label{3.12}&\gamma \in [2,\infty )\to \eta _1=-1, \eta _2=+1.
\end{align}
The cases $\gamma > -1/2$ for cosh and $\gamma < 1$ or $1<\ga <2$ for sinh are thus not subject to critical phenomena but may undergo phase transitions due to changes in the sign of $C_Q$. As a special case, we see that the \bh of the Eanti-Manti-D theory do not undergo any critical behavior in that their heat capacity never diverges.

The sinh solutions have another CP for $\ga \geq 2$, at extremality, where $C_Q$ diverges too but $T$ vanishes. This is denoted by $r_{\text{ext}}$ and is given by
\begin{equation}\label{3.12a}
    r_{\text{ext}}{}^2=2Q_{\text{ext}}{}^2/(1+\gamma),\quad (T_{\text{ext}}\equiv 0,\;Q_{\text{ext}}\text{: any}).
\end{equation}
In this case we will only consider the upper limit $r_+\to r_{\text{ext}}{}^+$ in the discussion of the critical behavior; in contrast with the point $r_c$ where both upper $r_+\to r_c{}^+$ and lower $r_+\to r_c{}^-$ limits will be considered. Combining $A_{\text{ext}}=Q_{\text{ext}}/r_{\text{ext}}$ with~\eqref{3.12a} we obtain
\begin{equation}\label{4.5}
    A_{\text{ext}}{}^2=(1+\gamma)/2\quad (\ga\geq 2)
\end{equation}

The physical case corresponds to $r_+ \geq r_-$ and thus the factor $[(1+\gamma ) r_+{}^2-2 \eta _2Q^2]/(1+\gamma )\propto r_+-r_-$, which appears in~\eqref{3.4}, \eqref{3.5}, and~\eqref{3.7}, is always positive or zero so that the sign of $C_Q$ depends only on the signs of the other factors in~\eqref{3.7}. In the $r_+$-$Q$ plane, $C_Q$ is positive between the two lines $L_1$ and $L_2$ defined by:
\begin{equation}\label{12bis}
\hspace{-1mm}L_1:\, Q=\sqrt{\Big|\frac{1+\ga}{2\ga}\Big|}r_+,\;  L_2:\, Q=\sqrt{\Big|\frac{1+\ga}{2(1+2\ga)}\Big|}\,r_+.
\end{equation}

Fig.~\ref{Fig1} and Fig.~\ref{Fig2} depict plots of $C_Q$ alone or of ($C_Q,M,T$) versus $A$ at constant $Q$ and plots of $Q$ versus $r_+$ (the lines $L_1$ and $L_2$) for the cosh solution of EManti-D with $\ga=-2$ and the sinh solution of EManti-D with $\ga=2$, respectively. For the sinh solution there is a line in the $r_+$-$Q$ plane that represents extremality and the \abh region is on the right of that line, while in the case of the cosh solution the \abh region is the whole first quadrant in the $r_+$-$Q$ plane. More other comments are provided in the captions of these figures and in Sect.~\ref{sec5}.

Three other formulas needed for Fig.~\ref{Fig2}, Fig.~\ref{Fig4} and subsequent discussions are the derivatives of the mass, temperature, and free energy $F_{\text{e}}\equiv M-TS$ with respect to $A$ at constant $Q$:
\begin{align}
&2Q(1+\ga)(\partial M/\partial A)_Q=-[(1+\gamma ) r_+{}^2-2 \eta_2\gamma  Q^2]\nn\\
\label{3.7bb}&Q(\partial T/\partial A)_Q=\frac{[(1+\ga)r_+{}^2-2\eta_2 Q^2]^{\ga-1}}{4\pi (1+\ga)^{\ga}r_+{}^{2\ga}}\times\\
&\quad \quad \quad \quad \quad \quad \quad [(1+\gamma ) r_+{}^2-2 \eta _2 (1+2 \gamma ) Q^2]\nn\\
&4Q(1+\ga)(\partial F_{\text{e}}/\partial A)_Q=-(1+\gamma ) r_+{}^2+2 \eta _2 (1+2 \gamma ) Q^2\nn.
\end{align}
Thus, as far as we keep the charge constant, the mass has its extremum at the point where $C_Q=0$ and the temperature and free energy have their extrema at the points where $C_Q$ diverges with $T_{\text{max}}=T_c$, which is the critical temperature corresponding to $r=r_c$, and
\begin{equation}\label{3.7cc}
    F_{\text{e}\,\text{min}}=\frac{Q_c}{\sqrt{2}}\sqrt{\frac{\eta_2(1+2\ga)}{1+\ga}}=\frac{Q_c}{2A_c}.
\end{equation}
This is shown explicitly in Fig.~\ref{Fig4} (a) where the plots $F_{\text{e}}(A)$ and $T(A)$ at constant $Q$ have been sketched using different vertical scales.

\section{Critical phenomena \label{sec5}}

In order to determine the properties of the transition at the CP for fluid-gas systems~\cite{book} it has become costume to fix $T$ and sketch the pressure $P$ in terms of the volume $V$; similar treatment is performed for magnetic systems~\cite{book}. In \abh thermodynamics, one proceeds the same way keeping $T$ constant and sketching the potential versus the charge or conversely~\cite{Wu} according to convenience and practicality [or, otherwise, sketching a given function of the charge $H(Q)$ versus another function of the potential $G(A)$].  In the following we will consider the $Q$-$A$ diagram at constant $T$ and treat separably the cases of the CP's ($A_c,Q_c,T_c$), which exists for both cosh and sinh solutions, and ($A_{\text{ext}},Q_{\text{ext}},T_{\text{ext}}$), which exists for sinh solutions only.

The main and mini plots of Fig.~\ref{Fig3} (a) are sketches of $Q(A)$ at constant temperature for the phantom cosh solution of EManti-D [Eq.~\eqref{3.9}] and the phantom sinh solution of EManti-D [Eq.~\eqref{3.12}], respectively. The phantom cosh solution of Eanti-MD [Eq.~\eqref{3.10}] has a plot $Q(A)$ [not shown in Fig.~\ref{Fig3} (a)] similar to that of the cosh solution of EManti-D [Eq.~\eqref{3.9}].

\subsection{cosh and sinh \BH: The nonextremality CP ($A_c,Q_c,T_c$)}

Substituting~\eqref{3.8} into~\eqref{3.5} or~\eqref{3.13} into the EOS~\eqref{3.6b} we determine the product $T_cQ_c$ by
\begin{equation}\label{3.14}
    T_cQ_c=\frac{2^{(2 \gamma -5)/2} }{\pi } \left(\frac{\gamma }{1+2 \gamma }\right)^{\gamma } \sqrt{\Big|\frac{1+\gamma }{1+2 \gamma }\Big|}.
\end{equation}
By analogy with fluid-gas or magnetic systems, we determine, for $T$ constant, the first and second derivatives $(\partial Q/\partial A)_T$ and $(\partial^2 Q/\partial A^2)_T$ on differentiating the EOS~\eqref{3.6b} then evaluate them at the CP using~\eqref{3.13}. Using the fact that
\begin{equation}\label{3.14b}
    F'(A)|_c = 0,\;F''(A)|_c = \frac{-\eta _2 A_c (2\gamma) ^{\gamma -1}}{\pi(1+\gamma )(1+2 \gamma )^{\gamma -2}}
\end{equation}
[where $F'(A)|_c=F'(A_c)$, $F''(A)|_c=F''(A_c)$] we obtain
\begin{equation}\label{3.15}
\Big(\frac{\partial Q}{\partial A}\Big)_T\Big|_c=0,\; T_c\Big(\frac{\partial^2 Q}{\partial A^2}\Big)_T\Big|_c=F''(A_c).
\end{equation}
We see that $(\partial^2 Q/\partial A^2)_T|_c$ is never zero. With only two equations, the EOS~\eqref{3.6} and the first equation in~\eqref{3.15}, we are thus unable to determine all the three thermodynamic variables ($A,Q,T$) at the CP in terms of the parameters of the problem ($\ga,\eta_1,\eta_2$): Only $A_c$ is known as function of ($\ga,\eta_1,\eta_2$). This is a new type of critical phenomena not encountered in classical thermodynamics~\cite{book3,book} nor is it, at least, in some AdS \bh~\cite{Wu}-\cite{AdS}. In some cases treated in~\cite{Wei}, as is the case of figure 6 (a) of~\cite{Wei}, where $\La =\La_{\text{c1}}=-2.5$, one encounters the same type of CP's. This type of transitions have been encountered in~\cite{Ma} too where critical behaviors of 3D black holes with torsion are dealt with.

In the case at hands, this means that for fixed ($\ga,\eta_1,\eta_2$), instead of having a point of inflection in the $Q$-$A$ diagram at the point ($A_c,Q_c$) where no coexistence should occur for $Q>Q_c$, rather we have an extremum value which is equal to $Q_c$ itself, as shown in Fig.~\ref{Fig3} (a), and (a set of CP's making up) a hyperbola in the $T$-$Q$ plane\footnote{Rigourously speaking, the hyperbola~\eqref{3.14} lies in the plane $A=A_c$ parallel to the $T$-$Q$ plane of the phase space ($A,Q,T$).} [Eq.~\eqref{3.14}] on which lie the points ($T_c,Q_c$) [or equivalently a line $L_2$ in the $r_+$-$Q$ plane given by~\eqref{12bis}]. By continuously lowering the value of $T_c$ at will, the new value of $Q_c$ increases, so we can observe coexistence at any value $Q$ of the charge.

\paragraph*{\textbf{Definition.}} \textit{First, because only the first derivative in~\eqref{3.15} is zero at the CP, we call this type of points first order CP's. Second, because the set of points ($T_c,Q_c$) make up a one-dimensional space in the phase space, we would rather call these points first order linear CP's in contrast with the case where one has one (or a set of discreet) CP(s) as in~\cite{Wei}. This notions generalize to cases where (1) all derivatives up to $n$ are zero: $(\partial^m Q/\partial A^m)_T|_c=0$ ($m\leq n$) and (2a) the CP's make up a (hyper)surface in the phase space: In that case we would have $n$\textsuperscript{th} order (hyper)surface CP's. Or (2b) the CP's do not make up any (hyper)surface, they would be called $n$\textsuperscript{th} order discreet CP's}.

This definition allows one to classify the CP's independently on the modern, or something else, classification of phase transitions. We will provide an example in Sect.~\ref{sec5b}.

The definition is certainly independent on ensembles. For fixed-charge ensembles where the charge of the \abh is held constant $Q=\text{const}$, we may, if we want, take $Q_c\equiv Q$, in this case all the other critical values ($A_c,T_c,S_c,\ldots$) are uniquely expressed in terms of ($\ga,\eta_1,\eta_2$) using the relevant equations.

Similarly, a $T$-$S$ diagram does not show any point of inflection, as in Fig.~\ref{Fig3} (b,c), but a turning point with a horizontal tangent line at ($S_c,T_c$), contrary to AdS \bh which have a point of inflection in the $T$-$S$ diagram~\cite{Sahay}.

Moreover, it is easy to show that, if $H\equiv H(Q)$ and $G\equiv G(A)$ are any functions of $Q$ and $A$, respectively, it is not possible to have at the same time $\dd H/\dd G|_c=0$ and $\dd ^2H/\dd G^2|_c=0$ with the requirement that $H$, $G$, and their first and second order derivatives be finite at the CP. In fact, if we assume that $H$, $G$, $H'$, $G'$, and $H''$ are finite at the CP, where the prime denotes the derivative with respect to the appropriate variable ($Q$ or $A$), using the first equation in~\eqref{3.15} we find
\begin{equation}\label{3.16}
  \Big(\frac{\partial H}{\partial G}\Big)_T\Big|_c=0, \quad \Big(\frac{\partial^2 H}{\partial G^2}\Big)_T\Big|_c=\Big(\frac{\partial^2 Q}{\partial A^2}\Big)_T\Big|_c\times\frac{1}{G'}\Big|_c
\end{equation}
so that by the second equation in~\eqref{3.15} and the requirement that $G'|_c$ be finite, we see that $\dd^2 H/\dd G^2|_c\neq 0$. This means that any thermodynamic variable transformation of the form $Q\to H(Q)$ and $A\to G(A)$, likely including Legendre transformations, would not lead to observe the point of inflection.

\paragraph*{\textbf{The critical exponents and scaling laws.}} The critical exponents characterize the phase transition near CP's. The latter are defined as in classical thermodynamics
\begin{align}
\label{4.1a}&Q-Q_c\sim |A-A_c|^{\de} & \text{ (for }\De T=0),\\
\label{4.1b}&C_Q\sim |Q-Q_c|^{-\bar{\phi}}& \text{ (for }\De T=0),\\
\label{4.1c}&S-S_c\sim |Q-Q_c|^{\psi}& \text{ (for }\De T=0),
\end{align}
\begin{align}
\label{4.1d}&A-A_c\sim |T-T_c|^{\bt}& \text{ (for }\De Q=0), \\
\label{4.1e}&C_Q\sim |T-T_c|^{-\al}& \text{ (for }\De Q=0), \\
\label{4.1f}&\kappa_T\sim |T-T_c|^{\ga '}& \text{ (for }\De Q=0),
\end{align}
where $\kappa_T\equiv (\partial Q/\partial A)_T/Q=F'(A)/F(A)$ is the factor of isothermal charge. In \abh thermodynamics, it has become custom to interchange the roles of extensive and intensive variables. In some references one chooses $\kappa_T\propto (\partial A/\partial Q)_T$~\cite{Wu} [and $\kappa_T\propto (\partial \Omega/\partial J)_T$ for rotating black holes~\cite{Mirza,Ma}] but one inserts a minus sign in front of $\ga '$ in~\eqref{4.1f}. Thus, whatever definition we give to $\kappa$, the value of $\ga '$ will be the same. The reason why some authors work with $\kappa_T\propto (\partial A/\partial Q)_T$ is that $(\partial A/\partial Q)_T$ diverges at criticality for a number of \abh solutions. This is going to be the case for some EMD \bh but not for all of them. Sticking to the definitions of classical thermodynamics, as some authors did~\cite{Banerjee:2011cz,Samanta,Sahay}, we have opted for the definition $\kappa_T\equiv (\partial Q/\partial A)_T/Q$ for convenience.

To ease the evaluation of the critical exponents, it is worth noting that the above-defined expressions that are valid for $\De T=0$ (respectively $\De Q=0$) can be combined together. For instance, combining~\eqref{4.1d} and~\eqref{4.1f} we obtain
\begin{equation}\label{4.1g}
    \kappa_T\sim |A-A_c|^{\ga '/\bt} \quad\text{(for }\De Q=0).
\end{equation}

By~\eqref{3.6b} and\eqref{3.14b} we have in the vicinity of the CP
\begin{equation}\label{4.1h}
    Q_c(T-T_c)+T_c(Q-Q_c)\sim [F''(A_c)/2](A-A_c)^2.
\end{equation}
If $\De T\equiv T-T_c=0$ (respectively $\De Q\equiv Q-Q_c=0$), this reduces to~\eqref{4.1a} [respectively to~\eqref{4.1d}] with $\de =2$ (respectively $\bt =1/2$).

For $\De T=0$, differentiating~\eqref{4.1h} with respect to $A$ we obtain
\begin{equation}\label{4.1i}
    \kappa_T\sim |A-A_c| \quad\text{(for }\De T=0).
\end{equation}
Now, since $\kappa_T=F'(A)/F(A)$ depends only on $A$, the series expansions in the r.h.s's of~\eqref{4.1g} and~\eqref{4.1i} must be identical leading to $\ga '=\bt=1/2$.

Combining~\eqref{4.1a} and~\eqref{4.1b}, from the one hand, and~\eqref{4.1d} and~\eqref{4.1e}, from the other hand, we obtain, respectively
\begin{align}
\label{4.2a}&C_Q\sim |A-A_c|^{-\de\bar{\phi}}\quad\text{(for }\De T=0),\\
\label{4.2b}&C_Q\sim |A-A_c|^{-\al/\bt}\quad\text{(for }\De Q=0).
\end{align}
With that said, the expression~\eqref{3.7} of $C_Q$ is easily brought to the form
\begin{equation}\label{4.2c}
    C_Q=Q^2\,\frac{\mathcal{C}(A)}{A-A_c}
\end{equation}
where $\mathcal{C}(A)$ is regular (finite) and nonzero at $A_c$. Thus if $\De Q=0$, $C_Q\sim |A-A_c|^{-1}$ yielding, on comparing with~\eqref{4.2b}, $\al =\bt=1/2$. If $\De T=0$, we have using~\eqref{4.1h}
\begin{equation}\label{4.2d}
Q\sim Q_c+[F''(A_c)/(2T_c)](A-A_c)^2\quad\text{(for }\De T=0)
\end{equation}
which along with~\eqref{4.2c} lead again to $C_Q\sim |A-A_c|^{-1}$. Using~\eqref{4.2a} we are finally led to $\bar{\phi}=1/\de=1/2$.

Combining~\eqref{4.1a} and~\eqref{4.1c} to obtain
\begin{equation}\label{4.3a}
    S-S_c\sim |A-A_c|^{\de\psi} \quad\text{(for }\De T=0).
\end{equation}
It is straightforward to bring the expression~\eqref{3.4} of $S$ to
\begin{equation}\label{4.3b}
    S=Q^2\mathcal{S}(A^2)
\end{equation}
where $\mathcal{S}(A^2)$ is regular (finite) at $A_c$. Now, for $\De T=0$ the expression of $Q$, given by~\eqref{4.2d}, and that of $\mathcal{S}$ being both regular at $A_c$, the Taylor series of $S$ at $A_c$ yields $S-S_c\sim |A-A_c|$, which is the same as~\eqref{4.3a} with $\psi=1/\de=1/2$.

The six critical exponents are tabulated in Table~\ref{Tab1}. They satisfy the following thermodynamic scaling laws
\begin{align}
&\alpha+2\beta+\gamma '=2,\;\alpha+\beta(\delta+1)=2 \nn\\
&(2-\alpha)(\delta\psi-1)+1=(1-\alpha)\delta  \nn\\
&\gamma '(\delta+1)=(2-\alpha)(\delta-1)\nn\\
\label{4.4}&\gamma '=\beta (\delta-1),\;\bar{\phi}+2\psi-\delta^{-1}=1.
\end{align}
\begin{table}[!htb]
\centering
\caption{\footnotesize Critical exponents corresponding to the CP ($A_c,Q_c,T_c$) given by Eqs.~\eqref{3.13} and~\eqref{3.14}. \label{Tab1}} \vspace*{0.3cm}
\begin{tabular}{|l|c|c|c|c|c|c|} %\hline
Symbol & $\alpha$ & $\beta$ & $\gamma '$
& $\delta$ & $\bar{\phi}$ & $\psi$  \\ \hline
Value & 1/2 & 1/2 & 1/2 & 2 & 1/2 & 1/2
\\ %\hline
\end{tabular}
\end{table}

\paragraph*{\textbf{On the continuous phase transition.}} Consider Fig.~\ref{Fig3} (a). The region $A>A_{\text{ext}}$ being excluded, we see that any horizontal line $Q=\text{const}<Q_c$ intersect the graph $Q(A)$ at two points, one of which lies on the unstable branch of the graph ($A<A_c$) and the other one lies on the stable branch ($A>A_c$). This shows the coexistence of the two (unstable and stable) phases for $Q<Q_c$; for $Q>Q_c$ there is no \aBH. This latter statement characterizes the phantom EM-antiD cosh and sinh \bh as well as the phantom Eanti-MD cosh holes\footnote{Recall that the phantom cosh solution of Eanti-MD [Eq.~\eqref{3.10}] has a plot $Q(A)$ (not shown in Fig.~\ref{Fig3} (a)) similar to that of the cosh solution of EManti-D [Eq.~\eqref{3.9}]}, in that, both unstable and stable phases either coexist or both do not exist. There is no configuration where only one phase exists. This is entirely different from what one has learnt in classical thermodynamics. To our knowledge there is no similar situation for other \BH.

The statement does not imply no coexistence for $Q>Q_c$. The product $T_cQ_c$ being fixed by~\eqref{3.14}, if we replot Fig.~\ref{Fig3} (a) for a value of the temperature $T_{c\,\text{new}}$ less than the selected value in Fig.~\ref{Fig3} (a) keeping the same parameters ($\ga,\eta_1,\eta_2$), the new value of the critical charge $Q_{c\,\text{new}}$ will exceed the old one and the crest in that figure rises, yielding coexistence for $Q_{c\,\text{new}}>Q>Q_c$. This is to say there is no true critical charge beyond which no coexistence occurs. This also distinguishes this type of phase transitions from ordinary ones~\cite{Sahay,Mo}.

Fig.~\ref{Fig3} (b,c) reveals the same coexistence state in the $T$-$S$ diagram where any horizontal line $T=\text{const}<T_c$ intersects the curve $T(S)$ at two points, one of which on the lower stable branch and the other one on the upper unstable branch. Here again we emphasize the special character of this transition by noticing that any reduction in the selected value $Q_c$ in Fig.~\ref{Fig3} (b,c) would yield an increase in the value of $T_c$, which would imply observation of coexistence at higher temperatures too.

A first conclusion we can draw from the above remarks is that stable and unstable heavily charged EManti-D cosh and sinh \bh and Eanti-MD cosh ones may undergo phase transitions at low temperatures only. As temperature rises, the two phases disappear along with the \aBH. Stable and unstable near-neutral ($Q\to 0$) EManti-D \bh may coexist at almost all temperatures and thus their phase transitions may occur at almost any temperature.

EManti-D and Eanti-MD \abh phase transitions being continuous and special they do however still fulfil the requirements of classical-thermodynamics phase transitions. Eqs~\eqref{3.7bb} and Fig.~\ref{Fig4} (a) show that the temperature has its local extreme values at the transition point (the CP), as is the case in classical thermodynamics~\cite{book} and in other \abh solutions~\cite{Banerjee:2011cz,Mo}. For canonical ensembles, with fixed charge, the relevant thermodynamic potential is the free energy $F_{\text{e}}$ depicted in Fig.~\ref{Fig4} (by analogy with classical thermodynamics, the free energy is the relevant thermodynamic potential for fixed-pressure ensembles). The free energy reaches its minimum value at the transition point as prescribed by classical thermodynamics. Fig.~\ref{Fig4} (b) is a parametric plot of $F_{\text{e}}(T)$ at constant $Q=Q_c$. The transition occurs at the lowest point of the $F_{\text{e}}(T)$ graph as one moves from the unstable upper branch to the stable lower one. A vertical line $T=\text{const}<T_c$ intersects the two branches of the $F_{\text{e}}(T)$ graph indicating again coexistence of unstable-stable phases.

\subsection{sinh \BH: The extremality CP ($A_{\text{ext}},Q_{\text{ext}},T_{\text{ext}}$)\label{sec5b}}

Besides the first order linear nonextremality CP ($A_c,Q_c,T_c$) given by Eqs.~\eqref{3.13} and~\eqref{3.14}, the sinh \bh have the extremality CP ($A_{\text{ext}},Q_{\text{ext}},T_{\text{ext}}\equiv 0$) given by Eqs.~\eqref{3.12a} and~\eqref{4.5} where $Q_{\text{ext}}$ and $r_{\text{ext}}$ remain arbitrary in that they are not determinable in terms of the parameters of the problem ($\ga,\eta_1,\eta_2$). In the mini plot of Fig.~\ref{Fig3} (a), we have taken $T=T_c\neq 0$; however, if we let $T_c\to T_{\text{ext}}=0$, the crest moves to infinity and the curve splits into the two vertical lines $A=0$ and $A=A_{\text{ext}}$, which are solutions of $F(A)=0$. On either line, the charge is undefined.

The set of CP's ($A_{\text{ext}},Q_{\text{ext}},T_{\text{ext}}$) make up the semi-line $A=\sqrt{(1+\gamma)/2}$ and $Q>0$ in the plane $T=0$ of the phase space ($A,Q,T$).

In terms of the mass $M$ of the hole we have
\begin{align*}
&\frac{2M}{1+\ga}=r_{\text{ext}}=\sqrt{\frac{2}{1+\ga}}\,Q_{\text{ext}}=r_-\\
&2M^2=(1+\ga)Q_{\text{ext}}{}^2.
\end{align*}
In an ensemble where $M$ is held constant, $Q_{\text{ext}}$ is fixed by the previous formula and the set of CP's reduces to a point. If $M$ fluctuates, then the CP's make up a segment of the above-mentioned semi-line bounded by two limiting values of $Q$.

As mentioned earlier in this work, we won't consider the case of extreme \bh since these are not stable and they may have nonthermal spectrum or ill-defined temperature~\cite{spec,spec2}. Rather, we consider the case of nonextremal sinh \abh approaching the extremality state defined by the CP ($A_{\text{ext}},Q_{\text{ext}},T_{\text{ext}}$). In this approach, thermodynamics is well defined and the spectrum of Hawking radiation is thermal.

\paragraph*{\textbf{The critical exponents and scaling laws.}} The evaluation of the critical exponents is carried out in absolutely the same way as that done in the previous section [Eqs.~\eqref{4.1a}-\eqref{4.3b}]. All one needs is to re-write the expressions of $F$, $C_Q$, and $S$ in a way to factor out $(A_{\text{ext}}-A)$
\begin{align}
\label{4.6a}&F(A)=\mathcal{F}(A)(A_{\text{ext}}-A)^{\ga}&\quad (\ga\geq 2)\\
\label{4.6b}&C_Q=Q^2\,\frac{\mathcal{\bar{C}}(A)}{(A_{\text{ext}}-A)^{\ga-1}}&\quad (\ga\geq 2)\\
\label{4.6c}&S=Q^2\,\frac{\mathcal{\bar{S}}(A)}{(A_{\text{ext}}-A)^{\ga-1}}&\quad (\ga\geq 2)
\end{align}
where $\mathcal{F}(A_{\text{ext}})\neq 0$, $\mathcal{\bar{C}}(A_{\text{ext}})\neq 0$, and $\mathcal{\bar{S}}(A_{\text{ext}})\neq 0$. We define $\De A\equiv A-A_{\text{ext}}<0$, $\De Q\equiv Q-Q_{\text{ext}}<0$, and $\De T\equiv T-T_{\text{ext}}=T>0$. The definitions of the critical exponents are given in Eqs.~\eqref{4.1a}-\eqref{4.1f} on replacing `$c$' by `$\text{ext}$' and Eq.~\eqref{4.1c} by
\begin{equation}\label{4.1cc}
    S\sim |Q-Q_{\text{ext}}|^{\psi}\quad \text{ (for }\De T=0)
\end{equation}
since $S_{\text{ext}}=\infty$.

Using~\eqref{4.6a}, Eq.~\eqref{4.1h} reads in the vicinity of the CP as
\begin{equation}\label{4.7}
Q\De T+T\De Q\sim (\De A)^{\ga},
\end{equation}
where $Q$ and $T$ are any values closer to $Q_{\text{ext}}$ and $T_{\text{ext}}=0$, respectively. In the limit $\De Q\to 0^-$, Eq.~\eqref{4.7} reduces to~\eqref{4.1d} with $\bt=1/\ga$. Similarly, in the limit $\De T\to 0^+$ (but $T\neq 0$), Eq.~\eqref{4.7} reduces to~\eqref{4.1a} with $\de=\ga$. With $F$ given by~\eqref{4.6a}, $\kappa_T=F'(A)/F(A)\sim |\De A|^{-1}$, which is the same as~\eqref{4.1g} with $\ga '=-\bt=-1/\ga$.

Using the expressions~\eqref{4.2a}, \eqref{4.2b}, and~\eqref{4.3a} (which reads $S\sim |A-A_c|^{\de\psi}$) along with~\eqref{4.6b} and~\eqref{4.6c} we evaluate the remaining critical exponents by: $\al =\bt (\ga -1)=(\ga -1)/\ga$, $\bar{\phi} =(\ga -1)/\de=(\ga -1)/\ga$, and $\psi=(1-\ga)/\de=(1-\ga)/\ga$. The six critical exponents are tabulated in Table~\ref{Tab2}. They satisfy the following thermodynamic scaling laws
\begin{align}
&\alpha+2\beta+\gamma '+1=2,\;\alpha+\beta(\delta+1)=2 \nn\\
&(2-\alpha)[\delta(\psi+1)-1]+1=(1-\alpha)\delta  \nn\\
&(\gamma '+1)(\delta+1)=(2-\alpha)(\delta-1)\nn\\
\label{4.8}&\gamma '+1=\beta (\delta-1),\;\bar{\phi}+2(\psi+1)-\delta^{-1}=1
\end{align}
which are the same as~\eqref{4.4} provided we transform $\ga '$ and $\psi$ by $\ga '\to\ga'+1$ and $\psi\to\psi+1$.

\begin{table}[!htb]
\centering
\caption{\footnotesize Critical exponents corresponding to the CP ($A_{\text{ext}},Q_{\text{ext}},T_{\text{ext}}$) given by Eqs.~\eqref{3.12a} and~\eqref{4.5}. \label{Tab2}} \vspace*{0.3cm}
\begin{tabular}{|l|c|c|c|c|c|c|} %\hline
Symbol & $\alpha$ & $\beta$ & $\gamma '$
& $\delta$ & $\bar{\phi}$ & $\psi$  \\ \hline
Value & $(\ga-1)/\ga$ & $1/\ga$ & $-1/\ga$ & $\ga$ & $(\ga-1)/\ga$ & $(1-\ga)/\ga$
\\ %\hline
\end{tabular}
\end{table}

\paragraph*{\textbf{On the phase transition.}} In Fig.~\ref{Fig2} (a), where $Q$ is held constant, the \abh region corresponds to $A<A_{\text{ext}}$. If we start from the left of $A_{\text{ext}}$ (near extremality) and we decrease the value of $A$, the entropy [Eq.~\eqref{4.6c}] decreases from infinity to a finite value. This corresponds to a transition from a disordered state to a more ordered one with lower value of the entropy. Such a transition of the symmetry is accompanied by a loss of mass\footnote{The loss of mass is justified analytically as follows. At criticality $A=A_{\text{ext}}$ where $A_{\text{ext}}$ is given by~\eqref{4.5}. Using this latter equation in~\eqref{F7}, or~\eqref{3.12a} in the first Eq.~\eqref{3.7bb}, yields $(\partial A/\partial M)_Q=(1+\ga)/[Q_{\text{ext}}(\ga -1)]>0$ at criticality ($\ga\geq 2$).} and an increase in the temperature, as shown in Fig.~\ref{Fig2} (a), as well as an increase in the radius of the horizon (since $Q=r_+A$ is held constant). With $Q$ held constant, any transition from extremality to nonextremality will cause emission of energy till the mass reaches its minimum value at the point where the heat capacity vanishes [see~\eqref{3.7}, \eqref{3.7bb}]. The nonextremal \abh thus formed will remain in an unstable or metastable state till $C_Q$ becomes positive where $M$ starts increasing. This is totally different from the case of \RN \abh where such a transition from extremality to nonextremality at constant $Q$ results in an increase in the mass, size of the hole, and temperature. The decrease in the entropy of the hole will be accompanied by an increase of the entropy of the system (hole plus environment).

Such a transition may take place if the near extremal \abh is isolated from its heat bath or environment, that is, if the hole is in a microcanonical state or in contact with the bath, in which case the hole is in a canonical ensemble. However, to reach the stable state, where $M$ increases with decreasing $A$ and $C_Q>0$, it is necessary to have the hole in a canonical state. Hence, if the charge of a sinh \abh of EManti-D, in a microcanonical ensemble near extremality, is held constant, the hole will evolve by emission of energy, which results in a decrease of its mass, to the final state of minimum mass and vanishing heat capacity.

This transition cannot be classified second order according to Ehrenfest scheme since the derivative of order one of the free energy, that is the entropy, diverges at the extremality CP. This expresses inaccuracy in Ehrenfest scheme. From this point of view, Hilfer developed a generalized scheme where the order of transition can be any positive real number~\cite{Hilfer,Hilfer2}. He argued that classical thermodynamics does not exclude the cases where the entropy diverges, as is the case at the extremality CP. It is clear that if the entropy diverges, the next derivatives of the free energy also diverge. According to Hilfer scheme such a transition is of order less than unity: $p<1$.

According to the definition given in the previous section, we now proceed to the classification of the CP ($A_{\text{ext}},Q_{\text{ext}},T_{\text{ext}}$). The $n$th order derivative of $Q$ with respect to $A$ at constant $T$, which is proportional to the $n$th order derivative of $F$ with respect to $A$, is not defined at criticality, so no classification scheme works. However, near criticality ($T\to T_{\text{ext}}=0$) the derivative reads
\begin{equation*}
T\Big(\frac{\partial^n Q}{\partial A^n}\Big)_T=(1+\ga-2A^2)^{\ga-n}P(A^2,\ga)
\end{equation*}
where $P$ is a polynomial in $A^2$ and $\ga$. Since $1+\ga-2A_{\text{ext}}{}^2=0$, one sees that all the derivatives up to order $n$ vanish at criticality if $n< \ga\leq n+1$. Thus, generically, the CP is an $n$\textsuperscript{th} order linear one if $n< \ga\leq n+1$. For specific values of $\ga$ satisfying $P(A_{\text{ext}}{}^2,\ga)=0$, the order of the CP may be higher than $n$.

\section{Thermodynamic fluctuations \label{secF}}

As is well known~\cite{book3}, there exist strong relationships between fluctuations of extensive thermodynamic entities and generalized heat capacities. Moreover, in both classical thermodynamics and \abh one, fluctuations and heat capacities diverge as criticality is approached. This shows that the study of thermodynamic fluctuations of a physical
system is intimately related to its stability and phase transitions.

Fluctuations of extensive thermodynamic parameters depend on how the system is perceived thermodynamically. This is because the evaluation of the fluctuation for a fluctuating extensive parameter is tied to the form of the probability distribution for that parameter, which is itself related to the ensemble in which the system makes part~\cite{book3,Kab1,Kab2}.

Different terminologies have been used to name the thermodynamic parameters: entropic intensive parameters in~\cite{book3} for intrinsic variables in~\cite{Kab1,Kab2}, and extensive parameters in~\cite{book3} for conjugate variables in~\cite{Kab1,Kab2}. But the formula for the second central moments, derived in~\cite{book3,Kab1,Kab2}, is just the same (compare Eq. (19.14) of~\cite{book3} with Eqs. (10) and (19) of~\cite{Kab1}). The reason for this different terminology is what we mentioned earlier, in \abh thermodynamics, it has become custom to interchange the roles of extensive and intensive variables. So, the terminology employed in~\cite{Kab1,Kab2} is more appropriate for \abh thermodynamics and likely for other fields of thermodynamics~\cite{new,newa,newb}.

If the system is envisaged in some thermodynamical ensemble (by ensemble we mean the set of constraints on the thermodynamic variables) and if $\Psi$ is the corresponding Massieu function~\cite{book3}, then the second central moments are equal to the second order derivatives of $\Psi$ (or to the first derivative of the conjugate variables) with respect to the intrinsic variables~\cite{Kab1}.

In the following we will consider separably the cases of the CP's ($A_c,Q_c,T_c$), which exists for both cosh and sinh solutions, and ($A_{\text{ext}},Q_{\text{ext}},T_{\text{ext}}$), which exists for sinh solutions only. We will be mostly interested in the moment of the mass fluctuation and some relative root mean square dispersions.

\subsection{cosh and sinh \BH: The nonextremality CP ($A_c,Q_c,T_c$)\label{secFa}}

Since $T_c\neq 0$, we assume that the \abh is in contact with a heat bath with which it exchanges heat only. The appropriate Massieu function is $\Psi=-\bt_T F_{\text{e}}=S-\bt_T M$ where $\bt_T=1/T$ and $Q$ are the intrinsic variables. The conjugate variables are $\{-M,-\eta_2\bt_T A\}$ since
\begin{equation}\label{F1}
\dd \Psi=-M\dd\bt_T-\eta_2\bt_T A\dd Q
\end{equation}
where we have used~\eqref{est}. The moment of the mass fluctuation is given by:
\begin{equation}\label{F2}
    \langle(\de M)^2\rangle=\Big(\frac{\partial^2 \Psi}{\partial \bt_T{}^2}\Big)_Q=\Big(\frac{\partial (-M)}{\partial \bt_T}\Big)_Q=T^2C_Q
\end{equation}
where $C_Q=(\partial M/\partial T)_Q$ has been used.

At criticality, the temperature $T$ of the heat bath approaches $T_c\neq 0$ and $C_Q$ approaches infinity as $|T-T_c|^{-\al}=|T-T_c|^{-1/2}$ [see Eq.~\eqref{4.1e} and Table~\ref{Tab1}]. As far as $T_c\neq 0$, we see that near criticality, the \abh may absorb from, or may release to, the heat bath (the Hawking bath) a huge amount of energy. If this happens, the thermodynamic description near criticality will breakdown. Such behavior has been noticed for \RN anti-de Sitter black holes too~\cite{Wu} and, to our knowledge, is common to all \BH.

The behavior near criticality has been described in~\cite{Kab1} as a mere change in stability, and not a phase transition. For instance, the \abh may completely evaporate if it releases all of its mass to the bath.

\subsection{sinh \BH: The extremality CP ($A_{\text{ext}},Q_{\text{ext}},T_{\text{ext}}$)\label{secFb}}

Consider again a canonical ensemble. With $T_{\text{ext}}= 0$ and $\al =(\ga-1)/\ga$ (Table~\ref{Tab2}), we have $\langle(\de M)^2\rangle \sim T^{(1+\ga)/\ga}$ as $T\to 0$ ($\ga\geq 2$). If it were possible to bring the temperature of the heat bath as close as possible to absolute zero, then the fluctuations in the mass would remain finite and vanish as $T\to 0$.

If the state of a heat bath with $T=0$ is not attainable by the third law of thermodynamics, then it is not possible to envisage these \bh in a canonical ensemble with $T=0$ held constant. Let us thus envisage the \abh in a microcanonical ensemble, that it, the hole is totally isolated from any environment. The appropriate Massieu function is its entropy $\Psi=S$. Using~\eqref{est} we have
\begin{equation}\label{F3}
    \dd \Psi=\bt_T\dd M-\eta_2\bt_T A\dd Q,\quad (\eta_2=1)
\end{equation}
so $\{M,Q\}$ and $\{\bt_T,-\eta_2\bt_T A\}$ are the intrinsic and conjugate variables, respectively. The moment in the fluctuation in the parameter $\bt_T$ is inversely proportional to that of the mass fluctuation: $\langle(\de \bt_T)^2\rangle =(\partial \bt_T/\partial M)_Q \sim T^{-(1+\ga)/\ga}$, which diverges as $T\to 0$. Since $\bt_T$ also diverges as $T\to 0$, it would be more appropriate to evaluate the relative root mean square dispersion: $\sqrt{\langle(\de \bt_T)^2\rangle}/\bt_T\sim T^{(\ga-1)/(2\ga)}$, which goes to zero as $T\to 0$ ($\ga\geq 2$) and there is no breakdown in thermodynamics.

We reach the same conclusion on calculating $\sqrt{\langle[\de (\bt_TA)]^2\rangle}/(\bt_TA)$. $\langle[\de (\bt_TA)]^2\rangle$ is evaluated by
\begin{multline}\label{F4}
    \langle[\de (\bt_TA)]^2\rangle=-\Big(\frac{\partial (\bt_T A)}{\partial Q}\Big)_M\\=-A\Big(\frac{\partial \bt_T}{\partial Q}\Big)_M-\bt_T\Big(\frac{\partial A}{\partial Q}\Big)_M.
\end{multline}
The Maxwell relation derived from~\eqref{F3} yields
\begin{align}
&\Big(\frac{\partial \bt_T}{\partial Q}\Big)_M=-\Big(\frac{\partial (\bt_T A)}{\partial M}\Big)_Q\nn\\
\label{F5}&\quad\quad\quad\quad =\frac{A}{T^2C_Q}-\bt_T\Big(\frac{\partial A}{\partial M}\Big)_Q.
\end{align}
To evaluate the two terms $(\partial A/\partial Q)_M$ in~\eqref{F4} and $(\partial A/\partial M)_Q$ in~\eqref{F5} we insert $r_+=Q/A$ [Eq.~\eqref{est1}] in~\eqref{3.3}. These two terms are given by [$(\partial A/\partial M)_Q$ has been evaluated in~\eqref{3.7bb}]
\begin{align}
\label{F6}&\Big(\frac{\partial A}{\partial Q}\Big)_M=-\frac{A}{Q}\,\frac{2\ga A^2+1+\ga}{2\ga A^2-1-\ga}\\
\label{F7}&\Big(\frac{\partial A}{\partial M}\Big)_Q=\frac{2A^2(1+\ga)}{Q(2\ga A^2-1-\ga)},
\end{align}
which remain finite at criticality [Eq.~\eqref{4.5}] but $\langle[\de (\bt_TA)]^2\rangle$ diverges as $T^{-(1+\ga)/\ga}$. It is now straightforward to see that $\langle[\de (\bt_TA)]^2\rangle/(\bt_TA)^2$ converges as $T\to 0$ ($\ga\geq 2$).

\section{Conclusion \label{secC}}

As we have seen the heat capacity of fixed-charge ensembles of EMD \bh may have both signs depending on the parameters of the theory. Leaving aside the normal \RN \aBH, we have shown that, among the other \bh of EMD theory, only the subclasses EManti-D and Eanti-MD of the cosh solutions and the subclass EManti-D of the sinh solutions exhibit critical phenomena if further the parameters ($\ga,\eta_1,\eta_2$) obey well defined equalities and inequalities.

These subclasses of \bh have a common nonextremality CP where the classical-thermodynamics scaling laws are preserved. This is however a first order linear CP, in that, a $Q$-$A$ diagram at constant $T$ does not show any point of inflection, leading to observe either coexistence of unstable and stable phases, below some `variable' critical charge, or absence of the two phases above it. As a consequence, heavily charged phantom \BH, of the above-mentioned subclasses, may undergo phase transitions at low temperatures only. In contrast, their counterpart near-neutral \bh may undergo phase transitions at almost any temperature.

The phase transitions at the nonextremality CP are classified continuous or second order by all existing classification schemes.

An extremality CP exists for the EManti-D sinh \bh where the scaling laws are satisfied provided two of the critical exponents are shifted by 1. This is a generically an $n$\textsuperscript{th} order linear CP if $n< \ga\leq n+1$ at which the extremality to nonextremality phase transition is of order less than unity according to Hilfer's generalized classification scheme. If the charge is held constant, the mass and entropy decrease while the temperature and radius of the horizon increase during the first metastable phase of this transition which ends as the mass reaches its minimum value. The transition may proceed to the more stable state with positive heat capacity and increasing mass if the \abh is not isolated; otherwise, the hole remains in the state of lower mass and vanishing heat capacity if isolated.

At the nonextremality CP, the fluctuations diverge if canonical ensembles are envisaged, which signals a breakdown of thermodynamics as criticality is approached. At the extremality CP, no breakdown of the laws has been noticed in both canonical and microcanonical ensembles, the only assumption we have made is that low temperatures are attainable in \abh thermodynamics.

%
%\section*{\large Appendix: ???}
%\renewcommand{\theequation}{A.\arabic{equation}}
%\setcounter{equation}{0}

%\newpage

\end{document}